\documentclass[conference]{IEEEtran}
\IEEEoverridecommandlockouts
\usepackage{url}
\usepackage{cite}
\usepackage{amsmath,amssymb,amsfonts}
\newcommand\tenpow[1]{\ensuremath{10^{#1}}}
\usepackage{algorithmic}
\usepackage{graphicx}
\usepackage{textcomp}
\usepackage{xcolor}
\usepackage{multirow}
\usepackage[font=normalsize]{caption}
\usepackage{booktabs}
\usepackage{makecell}
\usepackage[hidelinks]{hyperref}

\def\BibTeX{{\rm B\kern-.05em{\sc i\kern-.025em b}\kern-.08em
    T\kern-.1667em\lower.7ex\hbox{E}\kern-.125emX}}
\begin{document}

\title{Adaptive Control Attention Network for Underwater Acoustic Localization and Domain Adaptation\\
% {\footnotesize \textsuperscript{*}Note: Sub-titles are not captured for https://ieeexplore.ieee.org  and
% should not be used}
% \thanks{Thanks to the Office of Naval Research for funding (N00014-21-1-2790).}
% \thanks{}
}

\author{\IEEEauthorblockN{Quoc Thinh Vo, Joe Woods, Priontu Chowdhury, David K. Han}
\IEEEauthorblockA{\textit{Department of Electrical and Computer Engineering}, \textit{Drexel University}, Philadelphia, PA, USA \\
\{qv23, jw3897, pc833, dkh42\}@drexel.edu}
% \and
% \IEEEauthorblockN{2\textsuperscript{nd} Given Name Surname}
% \IEEEauthorblockA{\textit{Dept. of Electrical and Computer Engineering} \\
% \textit{Drexel University}\\
% Philadelphia, PA, USA \\
% 2@drexel.edu}
% \and
% \IEEEauthorblockN{3\textsuperscript{rd} Given Name Surname}
% \IEEEauthorblockA{\textit{Dept. of Electrical and Computer Engineering} \\
% \textit{Drexel University}\\
% Philadelphia, PA, USA \\
% 3@drexel.edu}
% \and
% \IEEEauthorblockN{4\textsuperscript{th} Given Name Surname}
% \IEEEauthorblockA{\textit{Dept. of Electrical and Computer Engineering} \\
% \textit{Drexel University}\\
% Philadelphia, PA, USA \\
% 4@drexel.edu}
}

\maketitle
\begin{abstract}
Localizing acoustic sound sources in the ocean is a challenging task due to the complex and dynamic nature of the environment. Factors such as high background noise, irregular underwater geometries, and varying acoustic properties make accurate localization difficult. To address these obstacles, we propose a multi-branch network architecture designed to accurately predict the distance between a moving acoustic source and a receiver, tested on real-world underwater signal arrays. The network leverages Convolutional Neural Networks (CNNs) for robust spatial feature extraction and integrates Conformers with self-attention mechanism to effectively capture temporal dependencies. Log-mel spectrogram and generalized cross-correlation with phase transform (GCC-PHAT) features are employed as input representations. To further enhance the model performance, we introduce an Adaptive Gain Control (AGC) layer, that adaptively adjusts the amplitude of input features, ensuring consistent energy levels across varying ranges, signal strengths, and noise conditions. We assess the model’s generalization capability by training it in one domain and testing it in a different domain, using only a limited amount of data from the test domain for fine-tuning. Our proposed method outperforms state-of-the-art (SOTA) approaches in similar settings, establishing new benchmarks for underwater sound localization.
\end{abstract}


\section{Introduction}

\begin{figure*}[htbp]
\centerline{\includegraphics[width=0.79\textwidth]{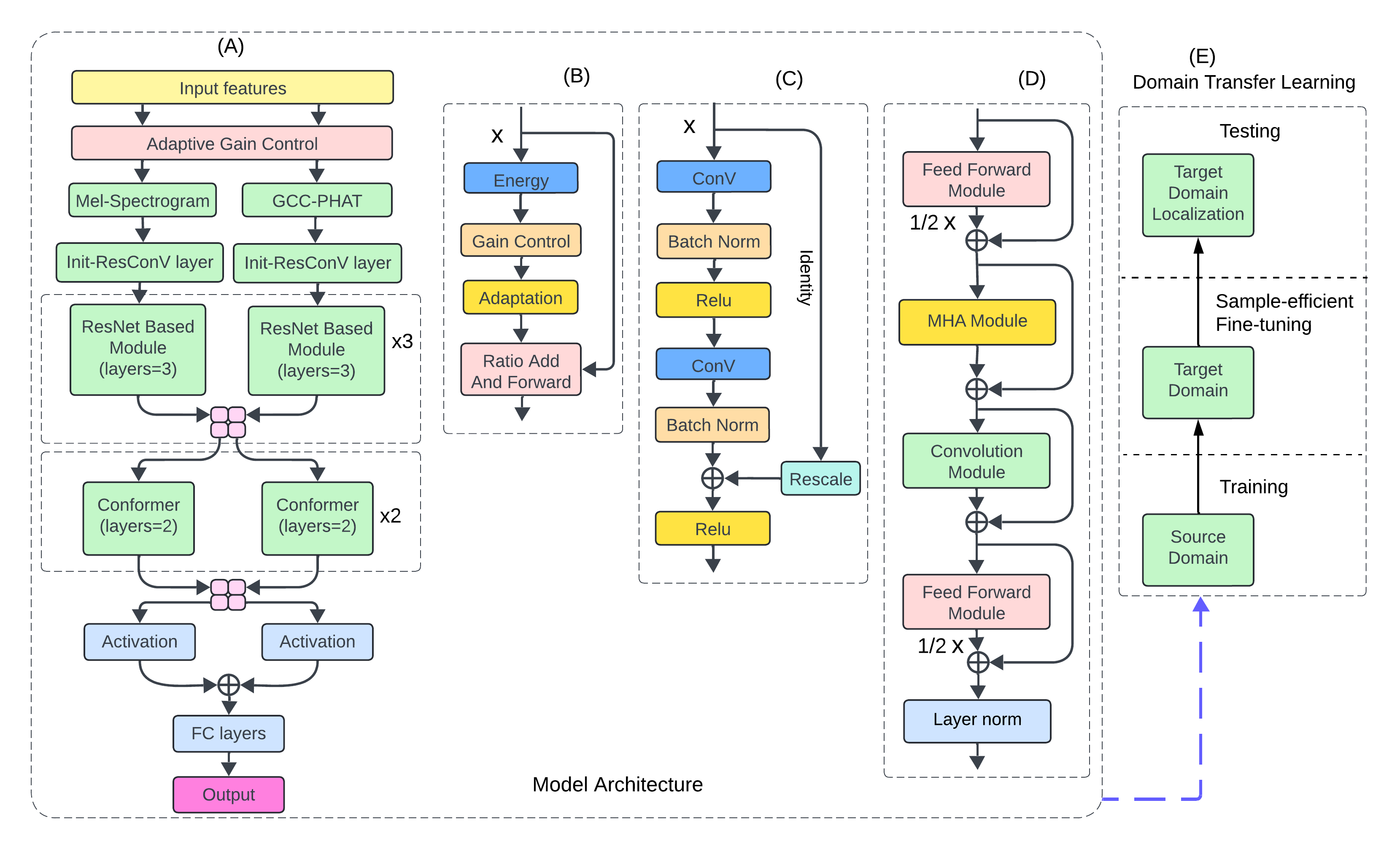}}
\caption{(A) The Proposed ACA-Net Architecture. (B) The Adaptive Gain Control Layer. (C) The Rescaled ResNet Based Module. (D) The Conformer Block. (E) The Domain Transfer Learning Framework.}
\label{architecture}
\end{figure*}

Sound source localization (SSL) is essential in underwater signal processing, with applications in communication, robotics, and surveillance \cite{Dong2024A, jurado2025underwater}. Varying water sound speed profiles, bottom sediment properties, surface dynamics, and background noise create complex propagation paths and interference, challenging accurate localization. 

Traditional SSL methods improve accuracy using physics-based techniques. Matched Field Processing (MFP) \cite{mfp_overview} uses known environment models to estimate source location, leveraging spatial patterns across sensor arrays for precise localization. It often relies on propagation models like KRAKEN \cite{porter1992kraken} to generate theoretical field replicas. Other traditional techniques include ESPRIT \cite{Kasthuri2021ESPRIT}, beamforming \cite{Wang2024Beamforming}, and Time Difference of Arrival \cite{Rezzouki2024TDOA}. These methods often struggle in complex acoustic conditions and depend on detailed environmental priors such as sound speed profiles, sediment properties, and bathymetry. 

% One MFP variant performs localization by matching the spatial covariance matrix (SCM) of the input data with replica vectors \( a \) representing candidate source locations at a particular frequency, defined as: 

% \textcolor{violet}{/ Was this equation in the introduction something we had before? Personally I'm not used to seeing mathematical notation so early on in a paper.}

% \begin{equation}
% \arg\max_{a} \left[ \mathbf{B}(a) = \mathbf{w}^H(a) \mathbf{P} \mathbf{w}(a) \right]
% \end{equation}

% where \( \mathbf{P} \) is the SCM of the input data, and  
% \( \mathbf{w}(a) \) represents the replica vector for a given location \( a \). Chen et al. \cite{chencnnresults} applied this to estimate source ranges in the SWellEx-96 dataset \cite{swellex96}, yielding a mean absolute error (MAE) of 1.73\,km (in the range 0.903–8.648 km). 

% \textcolor{violet}{/ I think the introduction still suffers from bouncing back and forth between generic SSL and underwater SSL, which Review #0244 mentioned as a problem.}

% Minimum Variance Distortionless Response (MVDR) \cite{kumar2016MVDR}
Recent advances in machine learning (ML) have significantly improved the performance of underwater signal and image processing tasks such as SSL, detection, and classification. CNNs have shown strong capabilities in sonar image recognition and underwater object detection, enabling automated identification and optimization tasks in complex underwater environments \cite{Guo2022UnderwaterCNN, Nga2024AutomatedCNN, Xu2024MainCNN}. Recurrent Neural Networks (RNNs), including hybrid architectures, have been effectively applied for time-series modeling and anomaly detection in acoustic sensing systems \cite{Davari2021RNN, Zhang2022ModulationRNN}. More recently, Transformer-based models have demonstrated robust performance in underwater acoustic classification and target recognition, even under noisy conditions, by capturing long-range temporal and spectral dependencies \cite{Feng2022Transformer, Vo2023MLSP-Transformer}. Additional information on the recent state of underwater acoustic signal processing for various tasks, including SSL, can be found in \cite{niu2023_advances}.

 The SWellEx-96 dataset \cite{swellex96} has been widely adopted as a benchmark for evaluating the performance of underwater SSL methods across a wide spectrum of approaches, from traditional signal processing techniques to deep learning models such as CNNs and Transformers. Notable approaches include Chen et al. \cite{chencnnresults} with the extension of CNNs enhancing performance in challenging environments and Chi et al. \cite{feast} addresses overfitting using fitting-based early stopping (FEAST).
% Wu et al. \cite{wu2021multi} introduced multi-task learning with Image Translation networks (MTIT), which addresses issues with single-task learning in indoor SSL but does not fully leverage contextual information.
Zhu et al. \cite{zhu_featurebased}, Wen et al. \cite{wensemi_siamese}, and Zhu et al. \cite{zhu2020semi} propose two-stage semi-supervised and self-supervised approaches to tackle data scarcity.  However, their solutions are computationally intensive due to additional training stages and underperform compared to SOTA methods. Moreover, current SOTA methods such as Wang et al. \cite{wangunetresults} with U-Net and He et al. \cite{he2024effective} with MLF-TransCNN, employ single-branch architectures for feature extraction. This design limits the ability to capture the complexity of multiple feature representations during extraction. In contrast, a multi-branch architecture allows independent extraction in each branch while leveraging correlations among distinct feature sets, thereby enhancing learning. Incorporating multiple features enriches input representations, improving accuracy, robustness, and adaptability, particularly in complex and noisy environments.

% Most SOTA methods, such as Wang et al. \cite{wangunetresults} with U-Net rely on single-branch forward-pass architectures, limiting their ability to capture the complexity of multi-feature input vectors. In contrast, multi-branch architectures enable independent feature extraction within branches, while still learning cross-correlation between features. He et al. \cite{he2024effective} employs multi-task learning for performance gains but requires high computational complexity, under-utilizes multi-branches, and relies on simulated data.

% This approach aids in discovering interdependencies between feature sets while enhancing classification and localization through the integration of more diverse input features.

% \textcolor{blue}{I still think that the sinlge task versus multitask is not describing what you are proposing.  It is rather that you are proposing a multibranch architecture vs single branch. With multibranch, it allows feature extractions that are focused on different set of things. Later the method allows cross attention among them for enhanced classification task. Please let me know your thoughts.}

To address the limitations of existing methods, we propose an Adaptive Control Attention Network (ACA-Net) featuring a multi-branch architecture where the branches share weights, facilitating a soft-sharing of parameters between them. Log-mel spectrogram and GCC-PHAT \cite{gcc-phat} features are used as inputs, with the former encoding spectral energy and the latter capturing inter-channel time delays. Additionally, we integrate an AGC layer, which dynamically adjusts the amplitude of input features, ensuring consistent energy levels to improve the model's performance and robustness. 
% This approach allows the model to use contextual information and capture interdependencies between features, enhancing its ability to learn diverse features in each branch, similar to how a cross-attention mechanism functions. 
We also use fine-tuning for domain adaptation demonstrating that the model can adapt to a target domain with high performance using only a few labeled samples. This reduces the reliance on extensive labels, addressing the data scarcity issues in underwater acoustics.
% This addresses the data scarcity issues in underwater acoustics.
% This reduces the reliance on extensive labels, addressing the data scarcity issues in underwater acoustics.
% \textcolor{blue}{I am not sure we should mention data augmentation here since there is not much of work we did in this topic.} \textcolor{red}{These techniques benefit from data augmentation methods such as spectrogram augmentation \cite{park2019specaugment} and synthetic room impulse responses \cite{roman2024spatial}.}
% \textcolor{blue}{Aren't all the methods passive here including the ones you mentioned earlier?} \textcolor{red}{In the passive localization field,}  notable
% \textcolor{blue}{Transition from the previous paragraph to this one is awkward. You need to tie the shortcomings of the methods you mentioned above with the solution you suggest. You need to step through the issues and the solutions you suggest for them.  In this way, you can say we propose these methods to address these key issues.}
% In this paper, we propose an Adaptive Gain Control ResNet-Conformer network for range estimation of a moving sound source. We also apply domain transfer techniques in a different domain experimental setup. 
% \textcolor{blue}{For each of these solutions, what issue is it addressing?}

Our proposed network achieves superior performance on the SWellEx-96 dataset, outperforming current SOTA solutions without the need for any additional data. The source-to-receiver distance (range) estimation task is formulated as a regression problem, which presents a greater challenge compared to the more commonly used classification approach \cite{niu2023_advances, jin_cassl}. The results highlight the effectiveness of our method, particularly in handling complex multi-channel environments, noisy underwater acoustic conditions, and domain adaptation.

% \begin{figure*}[htbp]
% \centerline{\includegraphics[width=\textwidth]{propose.png}}
% \caption{Proposed Network.}
% \label{fig}
% \end{figure*}

% \section{Overview of The Matched Field Processing}

% \textcolor{red}{Is SVP the only requirement? There area a slew of information required to make it work. Please list some of them.} 
\maketitle
\section{Proposed Methodology}
\subsection{Localization Model}
The proposed ACA-Net is designed based on the improved event-independent network multi-branch architecture \cite{cao2021improved, vo2024resnet}. In our work, we enhance the model by integrating ResNet-based blocks with rescaled residual connections and adding Conformers \cite{gulati2020conformer, peng2021conformer} to improve the capture of temporal dependencies. This implementation introduces a multi-scale attention mechanism to better capture local and global features across channels. To reduce feature space complexity and mitigate overfitting, we apply max-pooling and dropout after each block. The dual-branch model learns log-mel spectrogram features in one branch and GCC-PHAT features in the other. It also employs learnable shared weights in the middle \ensuremath{(2\times2)} windows. This approach enhances independent feature learning in each branch while still facilitating the discovery of interdependencies. The final layer is a multilayer perceptron of fully connected layers that outputs a single neuron predicting location across the time domain, treating range estimation as a regression problem. The base network size is $40,502,545$ parameters and its architecture is illustrated in Figure \ref{architecture}.

\begin{figure}
\centering
\includegraphics[width=8.5cm]{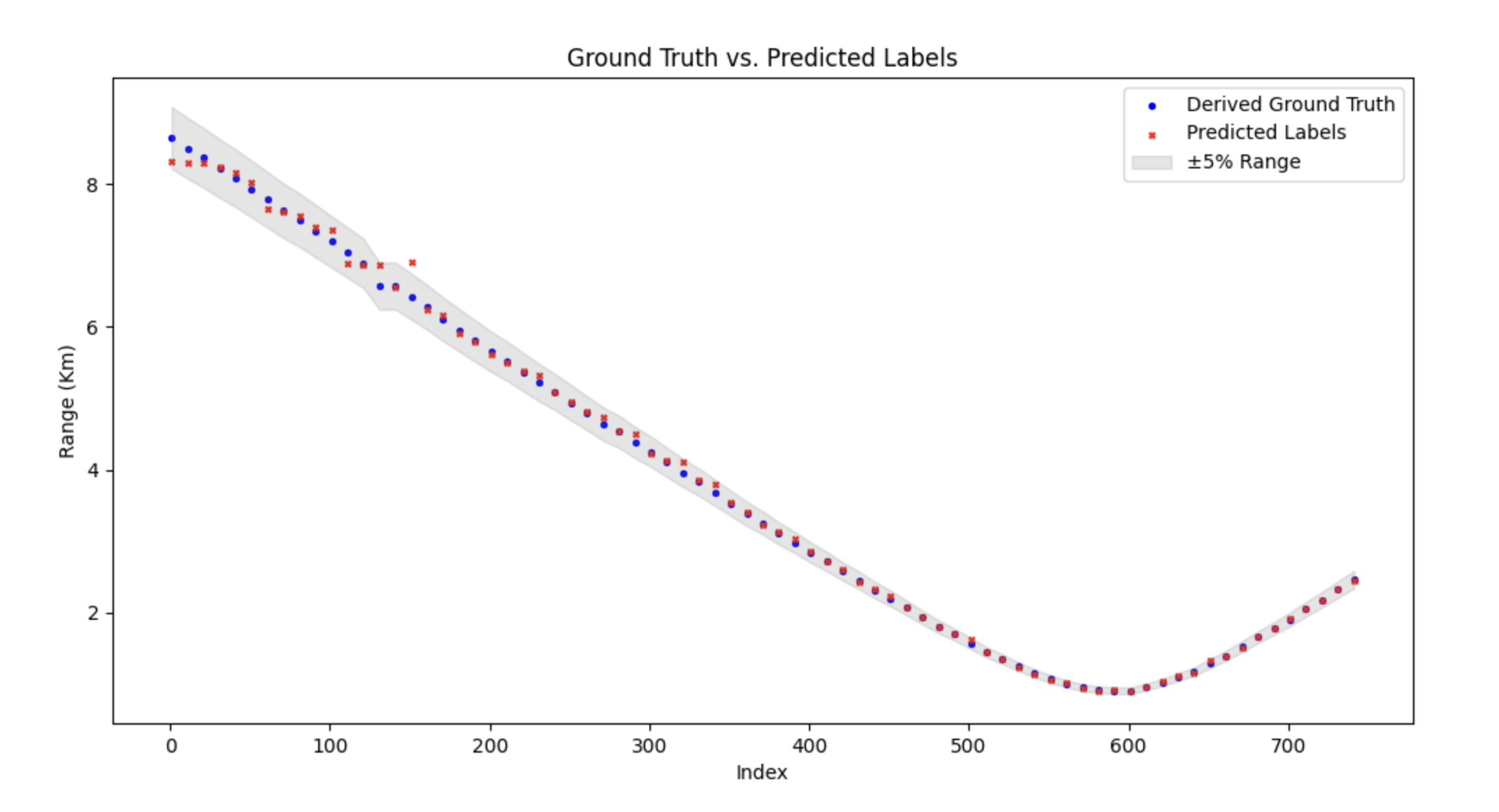}
% \caption{\textcolor{red}{Setup 1} \textcolor{blue}{In-domain S5 VLA }\textcolor{red}{- Every 10th point of 750 test samples is plotted - $P_{CL-5\%}$ = 98.62\% , MAE = 0.052}.}
% \caption{In-domain S5 VLA. The x-axis $"Index"$ represents test segment indices.\footnotemark}

\caption{In-domain S5 VLA. The x-axis \textquotedblleft Index\textquotedblright\ represents test segment indices.\protect\footnotemark[8]}
\label{fig:vla}
\end{figure}
% \textcolor{violet}{I think rather than saying what "Index" is, it would be better to change it to second since it basically represents time.}

% - $P_{CL-5\%}$ = 99.20\% , MAE = 0.045.}
% \textcolor{green}{I modified the way we present PCL and MAE to read a little differently than the other figures, let me know if preferred. -Joe}}

\begin{figure}
\centering
% \includegraphics[width=8.5cm]{true_vs_predicted_labels-56.89-0.081-new-left.png}
% \caption{Setup 2 - Event S5 VLA - Zero Shot - 0\% Labels Fine-tuning - 56.89\% ($P_{CL-5\%}$), 0.081 (MAE)}
\includegraphics[width=8.5cm]{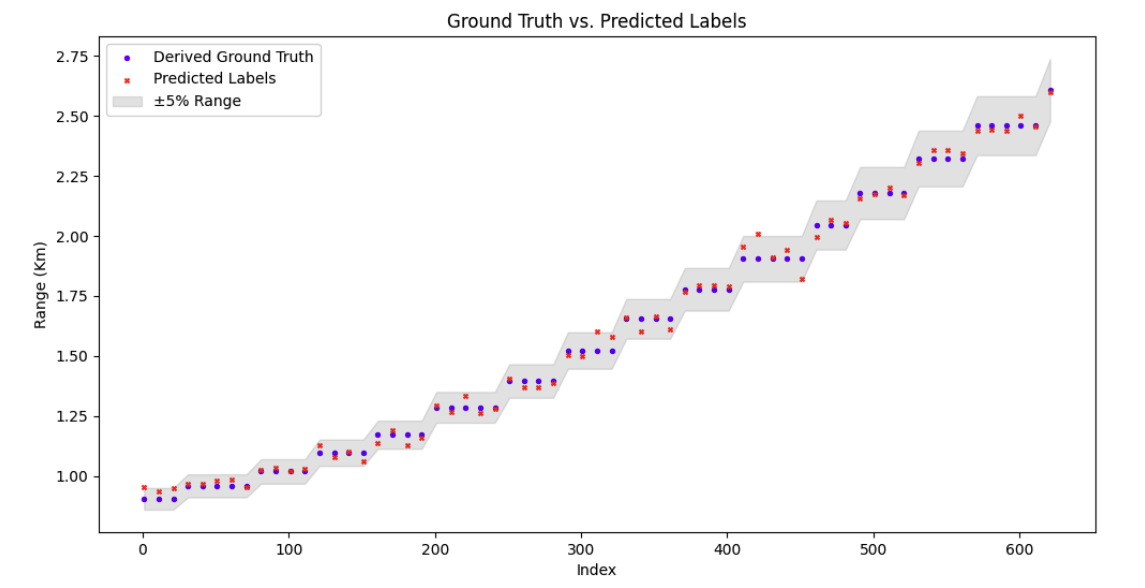}
\caption{Cross-domain Doppler Effect with 30\% labels fine-tuning. The x-axis \textquotedblleft Index\textquotedblright\ represents test segment indices.\protect\footnotemark[8]}
% - $P_{CL-5\%}$ = 96.67\%, MAE = 0.028.}
\label{fig:resultsleftright}
\end{figure}

\begin{figure}
\centering
\includegraphics[width=8cm]{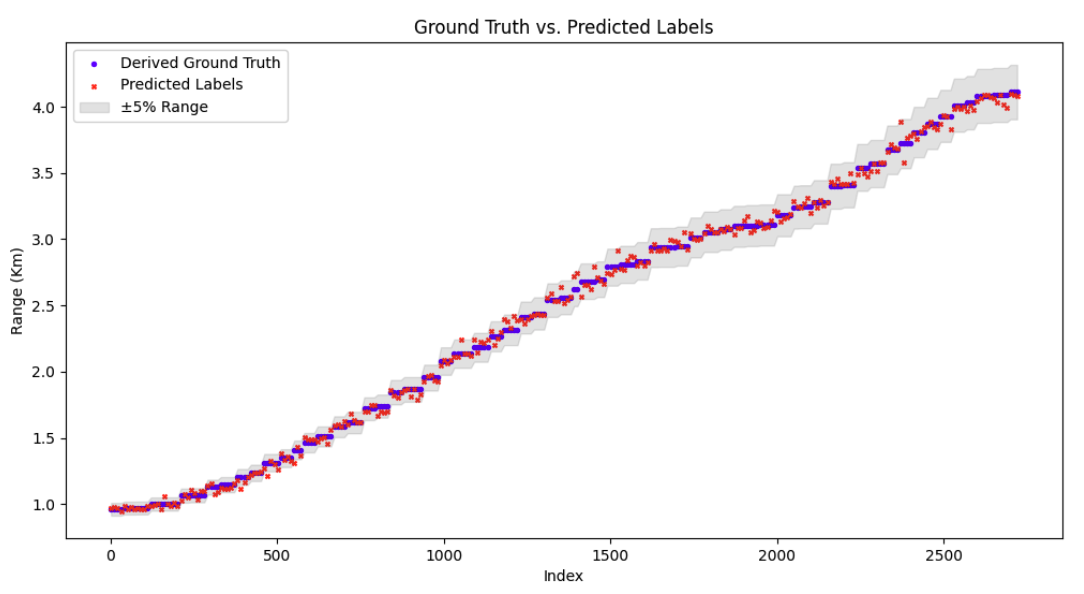}
\caption{Cross-domain S5 vs. S59 VLA with 30\% labels fine-tuning. The x-axis \textquotedblleft Index\textquotedblright\ represents test segment indices.\protect\footnotemark[8]}
% - $P_{CL-5\%}$ = 96.92\%, MAE = 0.032.}
\label{fig:s59and30per}
\end{figure}

\footnotetext[8]{ For better visualization, every \(10\)th indexed test segment is plotted.}

\subsection{Adaptive Gain Control (AGC) Layer}

The AGC layer is introduced to dynamically adjust the signal amplitude during processing the input features. Implemented as a dynamic control layer, the AGC adaptively maintains consistent energy levels across input features, which enhances the robustness of subsequent processing stages. The AGC calculates the input's energy and uses it to compute a gain factor that adjusts the input to a target energy level. To ensure numerical stability, a small constant, $\epsilon=\tenpow{-6}$, is added to the denominator of the function. The gain ensures smooth transitions and prevents abrupt changes that could disrupt learning, without requiring input normalization or rescaling, unlike many other methods \cite{wensemi_siamese, zhu_featurebased, zhu2020semi}. The energy per sample (\textit{E}) and the AGC output ($x_{adaptive}$) are computed as:
\begin{equation}
E = \frac{1}{N} \sum_{i=1}^{N} x_i^2
\end{equation}
\begin{equation}
x_{adaptive} = \left[ \left( \frac{E_{\text{target}}}{E + \epsilon} - 1 \right)\alpha  + 1 \right] x
\end{equation}

where N is the length of the input \textit{x}, \( E_{\text{target}} \) is the defined target energy, and \( \alpha \) is the adaptation rate.
We conducted a grid search for the best parameters: \(E_{\text{target}}\) = 1.0 and \( \alpha \) = 0.2.

% The AGC calculates the input signal’s energy and uses it to compute a gain factor that adjusts the signal to a target energy level. To ensure numerical stability and prevent division by zero, a small constant, $\epsilon=\tenpow{-6}$, is added to the denominator for the gain function. The gain is adaptively applied, controlled by an adaptation rate parameter that balances the gain adjustment with the original signal, ensuring smooth transitions and preventing abrupt changes that could disrupt learning.
% \begin{itemize}
%     \item \( E_{\text{target}} \) is the target energy,
%     \item \( E \) is the computed energy,
%     \item \( \alpha \) is the adaptation rate.
%     % \item \( \epsilon = 10^{-6} \) is a small constant to avoid division by zero.
% \end{itemize}

% The energy is calculated as:

% \begin{equation}
% E = \frac{1}{n} \sum_{i=1}^{n} x_i^2
% \end{equation}
% The AGC is particularly useful in underwater acoustic environments where signal amplitude varies significantly. By dynamical normalizing signal energy, the AGC enable the model to generalize better in tasks sensitive to amplitude variations. 
% \textcolor{blue}{Is it possible to add equations describing the process of AGC?}

% \subsection{Training}

All experiments are trained using supervised learning with backpropagation and mean squared error (MSE) loss, optimized with Adam (batch size 32, learning rate $10^{-4}$), on an NVIDIA RTX 3090 GPU. The MSE is defined as:

% \textcolor{blue}{If you can manage to squeeze in an equation here, it would be better.}

\begin{equation}
\label{equa1}
\text{MSE} = \frac{1}{N} \sum_{i=1}^{N} (y_i - \hat{y}_i)^2
\end{equation}

where N is the total segments, and $y_i$ and $\hat{y}_i$ are the ground truth and prediction at each segment index respectively.

% For Setup \ref{ssec:vlaset}, \ref{ssec:tlaset}, \ref{ssec:hla-northset}, \ref{ssec:hla-southset} 

\section{Data}

This study uses data from the S5 and S59 events of the SWellEx-96 experiment, which include Vertical Linear Array (VLA), Tilted Linear Array (TLA), Horizontal Linear Array (HLA) North, and HLA South. See \cite{swellex96} for more details.

Audio signals are segmented into non-overlapping one-second clips for all datasets. Each segment is labeled based on the closest "Duration" in minutes, as the original ground-truth range measurements are recorded once per minute. Log-mel spectrogram and GCC-PHAT features are extracted following the code and methods from \cite{vo2024resnet, krause2024sound}. We use a Short Time Fourier Transform (STFT) with a "Hann" window of length 40 samples and 50\% overlap to process spectrograms while capturing phase-difference cross-correlations between channel pairs to produce GCC-PHAT features.

\begin{table}
\caption{In-domain S5 VLA – ACA-Net ablation study}
\begin{center}
% \begin{flalign}
% \begin{tabular}{|c|c|c|c|}
% \begin{tabular}{lllllll}
\begin{tabular}{p{0.86cm} p{0.93cm} p{0.86cm} p{0.7cm} p{0.65cm} p{0.7cm} p{1.1cm}}

\toprule
\textbf{Mel-spectro-gram feature} & \textbf{ResNet Module} & 
\textbf{GCC-PHAT feature} & \textbf{Con-former} & \textbf{AGC Layer} & \textbf{MAE (km)} & \textbf{$\boldsymbol{P_{CL-5\%}}$ (\%)}
 \\
\midrule
% copy& More table copy$^{\mathrm{a}}$& &  \\
\checkmark & \checkmark &  & \checkmark & \checkmark & 0.131 & 77.31 \\
% \hline
\checkmark & \checkmark & \checkmark &  & \checkmark & 0.129 & 82.53 \\
% \hline
\checkmark & \checkmark & \checkmark & \checkmark &  & 0.064 & 96.93 \\
% \hline
\checkmark & \checkmark & \checkmark & \checkmark & \checkmark & \textbf{0.045} & \textbf{99.20} \\
 \bottomrule
 % \multicolumn{3}{1}{\textcolor{violet}{I think there is still a better way to visualize this ablation study, such as using a checkmark based system like found in the MLF-TransCNN Table 2}}
% \multicolumn{3}{l}{(\S) Experiments run with full network layers.} \\
% \multicolumn{3}{l}{(\dag) Experiments run with both mel-spectrogram and GCC-PHAT features.}
\end{tabular}
\label{ablation_study}
\end{center}
\end{table}

\begin{table}
\caption{In-domain S5 VLA - Range estimation comparison}
\centering
\begin{tabular}{llll}
\toprule
\textbf{} & \textbf{MAE (km)} & \textbf{MSE (km)}  &\textbf{$\boldsymbol{P_{CL-5\%}}$ (\%)} \\
\midrule
MFP \cite{chencnnresults} & 1.73 & --  & --\\
% \hline
CNN-r \cite{chencnnresults} & 1.40 & -- & --\\
% \hline
CPA-DDA-UNET \cite{wangunetresults} & 0.5976 & -- & --\\
% \hline
FEAST \cite{feast} & 0.5277 & -- & --\\
% \hline
Encoder-MLP \cite{zhu_featurebased} & -- & 0.22 & --\\
% \hline
Siamese-SSL \cite{wensemi_siamese} & -- & 0.1207 & --\\
% \hline
Time-Freq-CPC \cite{zhu2020semi} & -- & 0.11 & --\\
% \hline
MLF-TransCNN \cite{he2024effective} & 0.2718 & -- & --\\
% \hline
RA-CAE-SSL \cite{jin_cassl} & 0.0584 & -- & 65.50 \\
\midrule
\textbf{ACA-Net (Ours)} & \textbf{0.0452} & \textbf{0.00546} & \textbf{99.20} \\
 \bottomrule
% \multicolumn{3}{l}{\textbf{(*)} Previous SOTA performance for S5 event VLA dataset.}
\end{tabular}
\label{outside_comparison}
\end{table}

\section{Performance Metrics}
% \textcolor{green}{I think performance metrics would be better introduced before the results, that way the people looking at the figures know what PCL and MAE are before they get there. It may also be a good idea to separate the results from the figure captions? Though that might not really matter, it may feel more organized.}
% \textcolor{blue} {@Joe: I agreed with you on this - Quoc}
In our experiments, we establish performance benchmarks with mean absolute error (MAE) and Probability of Credible Localization ($P_{CL}$). The $P_{CL}$ metric measures the accuracy by considering predictions within a specified error limit. We evaluate the localization at a 5\% error. They are defined as:

\begin{equation}
P_{CL-5\%} = \frac{1}{N} \sum_{i=1}^{N} \mathbb{I}(i)
\end{equation}
\text{where:}
\begin{equation}
\mathbb{I}(i) =
\begin{cases} 
1, & \text{if } \frac{|y_i - \hat{y}_i|}{y_i} \times 100\% \leq 5\% \\
0, & \text{otherwise}
\end{cases} ,
\end{equation}

\begin{equation}
\text{MAE} = \frac{1}{N} \sum_{i=1}^{N} |y_i - \hat{y}_i|
\end{equation}

where N is the total segments, $y_{i}$ is the actual distance, and $\hat{y}_i$ is the predicted distance for the i-th index segment.
% \begin{figure}[htb]
% \centering
% \includegraphics[width=8.5cm]{true_vs_predicted_labels-100.png}
% \caption{Results Setup 1 - 98.62\% ($P_{CL-5\%}$) - 0.052 (MAE)}
% \label{fig:resultsall}
% \end{figure}

Additionally, Table~\ref{outside_comparison} reports the MSE (Eq.~\ref{equa1}) for comparison with methods that only provide benchmarks for MSE.

\section{Experimental Setup}

Our experiments include four in-domain setups for the VLA, TLA, HLA North, and HLA South arrays in S5 events, an ablation study, and two cross-domain setups (S5 VLA incoming vs. receding sound and S5 VLA vs. S59 VLA). In cross-domain experiments, we evaluate the network using zero-shot adaptation and fine-tuning with just 15\% and 30\% of target-domain data for sample-efficient domain transfer.

% \textcolor{blue}{[Maybe will be clearer if instead written "batch size of 32 and a learning rate of$ 10^{-4}$. -Priontu]}.

\subsection{In-Domain Experimental Setups}
\label{sec:in-domain}

% \textcolor{brown}{I think it may be better to refrain from numbering the setups and instead refer to them by their logical meaning (like "in-domain VLA"). I was also thinking maybe we can omit the figures for the in-domain TLA, HLA-North, and HLA-South. As of now we aren't analyzing them any differently than the VLA, and the VLA graph is representative of the behavior in the other ones. We can still include their results in the table and mention that the figures of the TLA and HLAs show similar trends to the VLA.}

For the in-domain S5 event VLA, TLA, HLA North, and HLA South, the data is evenly split into 6 folds: 4 for training, 1 for validation, and 1 for testing, ensuring balanced sample distribution. The process of constructing the 6-fold dataset is:
\begin{equation}
(x_i, y_i) \quad \forall i : fold = \mod(i,6)
\end{equation}

where \( fold \) is the fold number of the segment, \( x_i \) is the \( i \)th segment in the dataset, and \( y_i \) is the corresponding label.

The VLA and TLA datasets each contain 4,500 total segments, covering 75 minutes of recorded sound. Similarly, the HLA North and HLA South datasets each comprise 3,000 total segments, spanning 50 minutes.

\subsection{Cross-Domain Experimental Setups}
\label{sec:cross-domain}
In both cross-domain setups, we selected 0\%, 15\%, and 30\% random segments from the test domain in each fine-tuning experiments respectively and include the results in Table \ref{domain_adaptation_experiments}.

\subsubsection{Doppler Effect - S5 event VLA Approaching versus Receding Sound (Doppler)} \label{ssec:s5vla-cross}
Approaching sound data from 0 to 59 minutes is used for training as the source domain, while receding sound data from 60 to 75 minutes (900 segments) serves as the test set in the target domain. This setup emulates real-world Doppler effects and assesses the model's generalization capabilities. The maximum Doppler shift is given by:  
\begin{equation}
   \Delta f =  \frac{\pm 2.51}{1500} f_i \approx \pm 1.67 \times 10^{-3} f_i
\ 
\end{equation}

where \( f_i \) is the source frequency. This corresponds to an approximate shift in the pilot tones of $\pm 0.082$ to $\pm 0.67 \text{ Hz}$.

% To further evaluate our model's robustness, we designed a second experimental setup to introduce a domain adaptation challenge. In this setup, we used the S5 event VLA dataset, with approaching sound as the source domain data for training and receding sound as the target domain data for testing. This configuration simulates real-world Doppler effects and offers insights into the model's generalization capabilities. Specifically, data from 0 to 59 minutes of approaching sound served as the training set, while data from 60 to 75 minutes of receding sound was used for testing.}

% \textcolor{brown}{I'm confused how the Fine-Tuning described in the caption relates to the description given here. Is it all data 0-59 + 30\% of data 60-75 that is used as training? How as that 30\% selected? Also need to quantitatively show the Doppler Effect to prove that this is a different domain}
% \textcolor{blue}{The 30\% is selected randomly and would be from the test set to make sense. 
% I described above that: Approaching sound data from 0 to 59 minutes was used for training as the source domain, while receding sound data from 60 to 75 minutes served as the test set as target domain. }

% \textcolor{red}{Plot Figure 4 with fewer points so that the ground truth and the prediction points can be distinguished.}
% The data split for this experiment is illustrated in Figure \ref{fig:trainleft}.

% \begin{figure}[htb]
% \centering
% \includegraphics[width=8.5cm]{trainleft-testright.png}
% \caption{Second Setup: Different Domain Experiment Data split.}
% \label{fig:trainleft}
% \end{figure}

\subsubsection{S5 VLA versus S59 VLA (S5 vs. S59)}
\label{ssec:s5vss59}
The model is trained on the S5 event VLA dataset and tested on the S59 event VLA dataset. Event S59 also involves a source towed along an isobath but includes an interferer nearby. The S59 event includes 65 minutes of audio, totaling 3900 one-second segments. This is useful for examining how a loud interferer affects the localization of a quiet target, presenting a greater challenge to the model in the target domain.

% \subsection{Ablation Study}

\section{Results}

\begin{table}
\caption{ACA-Net - In-domain setup performances}
\begin{center}
\begin{tabular}{lll}
\toprule
 &{\textbf{MAE (km)}} & \textbf{$\boldsymbol{P_{CL-5\%}}$(\%)}\\
\midrule
S5 event VLA & 0.045 & 99.20 \\
% \hline
S5 event TLA & 0.044 & 98.67 \\
% \hline
S5 event HLA North & 0.024 & 84.20  \\
% \hline
S5 event HLA South & 0.037 & 81.80 \\
\bottomrule
\multicolumn{3}{l}{($\ast$) All of our established benchmarks are SOTA performances.}
\end{tabular}
\label{in_domain_experiments}
\end{center}
\end{table}

\begin{table}
\caption{ACA-Net - Cross-domain setup performances}
\begin{centering}
% \begin{flalign}
% \begin{tabular}{|c|c|c|c|}
\begin{tabular}{llll}
\toprule
 \textbf{Setup} & \textbf{Data Used (\%)} & {\textbf{MAE (km)}} & \textbf{$\boldsymbol{P_{CL-5\%}}$(\%)}
 \\
% \cline{2-4} 
% \textbf{Head} & \textbf{\textit{Table column subhead}}& \textbf{\textit{Subhead}}& \textbf{\textit{Subhead}} \\
\midrule
% \multicolumn{3}{l}{\textbf{Doppler Effect}} \\
\multirow{3}{*}{{\textbf{Doppler}}} &
% copy& More table copy$^{\mathrm{a}}$& &  \\
Zero-shot 0\% (\S)  & 0.078 & 59.11  \\
% \hline
& Fine-tuned 15\% (\S) & 0.032 & 92.94 \\
% \hline
& \textbf{Fine-tuned 30\%} (\S) & \textbf{0.025} & \textbf{98.57}\\
& Trained-scratch 15\% (\dag) & 0.113 & 44.44 \\
% \hline
& Trained-scratch 30\% (\dag) & 0.057 & 77.46 \\
\midrule
% \multicolumn{3}{l}{\textbf{S5 vs. S59 VLA}} \\

\multirow{3}{*}{{\textbf{S5 vs. S59}}} & Zero-shot 0\% (\S) & 0.441 & 15.21 \\
% \hline
& Fine-tuned 15\% (\S)  & 0.070 & 81.06\\
% \hline
& \textbf{Fine-tuned 30\%} (\S) & \textbf{0.032} & \textbf{96.92} \\
& Trained-scratch 15\% (\dag) & 0.128 & 54.54 \\
% \hline
& Trained-scratch 30\% (\dag) & 0.061 & 89.27 \\
\bottomrule
\multicolumn{4}{l}{(\S) Source domain pre-trained; fine-tuned with test domain data (\%).} \\
\multicolumn{4}{l}{(\dag) No pre-training; trained with test domain data (\%) from scratch.}
\end{tabular}
\label{domain_adaptation_experiments}
\end{centering}
\end{table}

We conduct an ablation study to evaluate the effectiveness of key layer components and the impact of different selective feature sets as shown in Table \ref{ablation_study}. It allows us to learn that the Conformer layer is the greatest contribution to our model, followed by GCC-PHAT feature and the AGC layer.

Table~\ref{outside_comparison} compares our model with recent methods, showing that ACA-Net achieves SOTA performance on the SwellEx-96 S5 event VLA, as illustrated in Figure~\ref{fig:vla}. Notably, the RA-CAE-SSL method \cite{jin_cassl} applies selective band-pass filters based on known frequencies to reduce noise and frames range estimation as a 100-class classification task within a narrow test range (0.903–2.608 km). Although it outperforms other SOTA methods, its dependence on prior frequency knowledge and fixed range bins limits scalability to unknown environments and makes it unsuitable for the regression task in this study. Compared to the more relevent current SOTA regression method MLF-TransCNN \cite{he2024effective}, our network achieves a significantly lower MAE of 0.0452 km versus 0.2718 km.

Similar trends, such as lower performance on long-range estimation, are observed for the TLA and HLAs, but figures are omitted due to the page limit. Table \ref{in_domain_experiments} displays results from our various in-domain experiments. We are among the first to conduct experiments on the S5 TLA, HLA North and South. ACA-Net demonstrates robust performance across various multi-channel array configurations. The HLAs arrays exhibit lower $P_{CL-5\%}$ scores, indicating reduced accuracy due to fewer data samples (50 minutes) compared to the VLA and TLA arrays (75 minutes). Despite this, they achieve slightly better MAE, likely because their recordings cover narrower range intervals in those 50 minutes (See \cite{swellex96} for more details). Meanwhile, the TLA and VLA, with equal durations and identical range coverage, show consistently comparable performance as reported in Table~\ref{in_domain_experiments}.

Additionally, the results in Table~\ref{domain_adaptation_experiments} highlight the effectiveness of source-domain pre-training for cross-domain adaptation. As test-domain data decreases, the benefits of pre-training become more pronounced, especially in the Doppler Effect experiment where test data is more limited. The source-pretrained network achieves zero-shot performance only slightly below that of training from scratch with 30\% of the target domain data. Similarly, in the S5 vs. S59 setting, fine-tuning the source-pretrained model with only 15\% of test domain data achieves performance on par with training from scratch with 30\% of the data. Table \ref{domain_adaptation_experiments} and Figures~\ref{fig:resultsleftright} and \ref{fig:s59and30per} show that our model is robust to the Doppler Effect and loud acoustic interference, suggesting generalization capability in domain adaptation with minimal training or fine-tuning.

% \textcolor{red}{This figure is not referenced properly! I'm not sure how we want to organize the results section, so might want to discuss this before worrying too much.} 

\section{Conclusions}

Our proposed ACA-Net outperforms SOTA methods on the S5 event VLA, TLA, HLA North, and HLA South dataset, achieving strong cross-domain adaptation with minimal fine-tuning in the S5 VLA vs. S59 VLA and Doppler Effect experiments, suggesting a strong generalization. It shows the benefits of using multiple features via a multi-branch architecture and the ability to adapt to a new array dataset efficiently. While the network achieves superior performance with a range of arrays and events, it has not yet been evaluated under a wider variety of oceanic conditions. Future research will focus on creating a dynamic feature extraction layer to process signals from various arrays, allowing localization independent of array structure, and will involve testing the network on additional underwater datasets. Additional acoustic information may also be used to enhance training and reduce data requirements through a more physics-informed ML approach.

% \textcolor{brown}{May also want to mention some limitations of our work}
% \textcolor{blue}{What is the SOTA compared here?}

% \textcolor{blue}{Is the plan to develop better branches for evaluation or for better performance?}

\section*{Acknowledgment}
The work on this paper was supported by the Office of Naval Research (Grant No. N00014-21-1-2790).

\bibliographystyle{IEEEtran}   % IEEE style referencing
\bibliography{references}

\vspace{12pt}

\end{document}